\documentclass[aps,prb,reprint,superscriptaddress]{revtex4-1}

\usepackage{graphicx}
\usepackage{dcolumn}
\usepackage{amsmath}
\usepackage{textcomp}
\usepackage{xcolor}
\usepackage{hyperref}

\begin{document}

\title{Phonon-trapping enhanced energy resolution in superconducting single photon detectors}

\author{Pieter J. de Visser}
\email{p.j.de.visser@sron.nl}
\affiliation{SRON Netherlands Institute for Space Research, Sorbonnelaan 2, 3584 CA Utrecht, The Netherlands}

\author{Steven A. H. de Rooij}
\affiliation{SRON Netherlands Institute for Space Research, Sorbonnelaan 2, 3584 CA Utrecht, The Netherlands}
\affiliation{Faculty of Electrical Engineering, Mathematics and Computer Science, Delft University of Technology, Mekelweg 4, 2628 CD Delft, The Netherlands}

\author{Vignesh Murugesan}
\affiliation{SRON Netherlands Institute for Space Research, Sorbonnelaan 2, 3584 CA Utrecht, The Netherlands}

\author{David J. Thoen}
\affiliation{Faculty of Electrical Engineering, Mathematics and Computer Science, Delft University of Technology, Mekelweg 4, 2628 CD Delft, The Netherlands}
\affiliation{Delft University of Technology, Kavli Institute of NanoScience, Faculty of Applied Sciences, Delft, The Netherlands}

\author{Jochem J. A. Baselmans}
\affiliation{SRON Netherlands Institute for Space Research, Sorbonnelaan 2, 3584 CA Utrecht, The Netherlands}
\affiliation{Faculty of Electrical Engineering, Mathematics and Computer Science, Delft University of Technology, Mekelweg 4, 2628 CD Delft, The Netherlands}

\date{\today}

\begin{abstract}

A noiseless, photon counting detector, which resolves the energy of each photon, could radically change astronomy, biophysics and quantum optics. Superconducting detectors promise an intrinsic resolving power at visible wavelengths of $R=E/\delta E\approx100$ due to their low excitation energy. We study superconducting energy-resolving Microwave Kinetic Inductance Detectors (MKIDs), which hold particular promise for larger cameras. A visible/near-infrared photon absorbed in the superconductor creates a few thousand quasiparticles through several stages of electron-phonon interaction. Here we demonstrate experimentally that the resolving power of MKIDs at visible to near-infrared wavelengths is limited by the loss of hot phonons during this process. We measure the resolving power of our aluminum-based detector as a function of photon energy using four lasers with wavelengths between $1545-402$ nm. For detectors on thick SiN/Si and sapphire substrates the resolving power is limited to $10-21$ for the respective wavelengths, consistent with the loss of hot phonons. When we suspend the sensitive part of the detector on a 110 nm thick SiN membrane, the measured resolving power improves to $19-52$ respectively. The improvement is equivalent to a factor $8\pm2$ stronger phonon trapping on the membrane, which is consistent with a geometrical phonon propagation model for these hot phonons. We discuss a route towards the Fano limit by phonon engineering.

\end{abstract}

\maketitle

\section{Introduction}\label{sec:introduction}
A camera in which each pixel can count individual photons with high efficiency and which simultaneously resolves each photon's colour is a dream for many (bio)physicists, astronomers and anybody working in low light applications. In semiconductor based detectors, each visible or near-infrared photon creates only one or a few excitations due to the large bandgap. Therefore energy information about absorbed photons is lost. Dark excitations are indistinguishable from photon excitations, which causes the well-known dark current and read noise that limit the detector sensitivity. In a superconductor, the energy needed to create excitations can be orders of magnitude less than in a semiconductor. A single visible photon can break a few thousand Cooper pairs. A measurement of the number of broken pairs is therefore a direct measure of the photon energy. Microwave kinetic inductance detectors (MKIDs) \cite{pday2003,jzmuidzinas2012} measure the number of quasiparticles excited in the superconductor through a change in the complex conductivity. The change in kinetic inductance and resistance of the superconductor are measured with a microwave resonator. Microwave resonators can be naturally multiplexed into large arrays. Several visible/near-infrared MKID instruments are deployed at telescopes, with up to 20k pixels \cite{bmazin2012,smeeker2018,awalter2018}. Other energy-resolving detectors for these wavelengths are Transition Edge Sensors (TES) and Superconducting Tunnel Junction detectors (STJ), which are harder to mutiplex into arrays, although small arrays have been demonstrated in astronomical \cite{rromani2001,pverhoeve2006} and biological \cite{kniwa2017} applications.

One of the main characteristics of photon counting MKIDs, the energy resolving power, is poorly understood. The resolving power can be limited by different mechanisms, such as: the signal to noise, quasiparticle diffusion, non-uniformity of the microwave current and phonon losses. Ultimately, the resolving power is limited by the statistical variation in the number of quasiparticles generated per photon. This limit arises because the initial photon energy (a few eV) is downconverted into quasiparticles with a typical energy of $0.1-0.2$ meV through a number of electron-electron and electron-phonon interaction steps. When all photon energy is kept inside the detector during this pre-detection stage, the resolving power is described by Fano statistics \cite{mkurakado1982,nrando1992}, and given by
\begin{equation}
R_{Fano} = \frac{1}{2\sqrt{2 \ln2}}\sqrt{\frac{\eta_{pb}^{max}E}{\Delta F}},
\label{eq:Fano}
\end{equation}
with $E$ the photon energy, $\Delta=1.76k_BT_c$ the gap energy of the superconductor, and $T_c$ the critical temperature. $\eta_{pb}$ expresses that not all energy can be converted to new quasiparticles. Even when all photon energy stays inside the superconductor during downconversion, $\eta_{pb} = \eta_{pb}^{max}= 0.59$, which is equivalent to an average energy of 1.7$\Delta$ per generated quasiparticle\cite{akozorezov2000}. $F \approx 0.2$ is the Fano factor for superconductors \cite{mkurakado1982,nrando1992}. For aluminum with $T_c = 1.25$ K, $R_{Fano} = 47-92$ for wavelengths of $1550-400$ nm respectively, and would be $R_{Fano} = 80-160$ for $T_c = 0.4$ K. Throughout this paper we will use the term resolving power for $R=E/\delta E$, with $\delta E$ the full-width-half-maximum (FWHM) energy resolution, expressed in energy units.

Energy resolution measurements on MKIDs have been reported for different materials. For TiN (with $T_c=0.8$ K), $R\approx 10$ at 400 nm \cite{bmazin2012,bmazin2013}. For PtSi, with $T_c=0.9$ K, $R=5.8-8.1$ at wavelengths of $1310-808$ nm (a best $\delta E$ = 0.16 eV)\cite{pszypryt2017}. For Hf with $T_c=0.4$ K, $R=6.8-9.4$ (best $\delta E$ = 0.14 eV) for the same wavelengths \cite{nzobrist2019b}. In devices optimised for photon-number resolution using TiN/Ti/TiN trilayers  $R=3.6$ at 1550 nm\cite{wguo2017}. All of these results are partially due to noise limitations and partially due to unknown reasons. Care was taken to not be limited by current non-uniformity\cite{bmazin2012,nzobrist2019}. The best resolving powers of $R=9.6-17$ for wavelengths of $1310-406$ nm (best $\delta E$ = 0.10 eV) were recently reported using a parametric amplifier to reduce the amplifier noise \cite{nzobrist2019,nzobrist2021}. In Ref. \onlinecite{nzobrist2019}, the signal to noise limit was $R_{SN}=24$ at 1120 nm, which clearly shows that another mechanism dominates $R$, possibly phonon losses. These numbers show that the Fano limit is still far out of reach for MKIDs. On top of that, to study the intrinsic resolving power limits, a much higher signal to noise is desirable. 

TES detectors have shown resolving powers of $11-23$ at $780-350$ nm \cite{bcabrera1998} (a constant $\delta E=$ 0.15 eV) and 7 at 1570 nm \cite{llolli2013}. TES have proven very powerful in counting the number of photons in pulses \cite{llolli2013,drosenberg2005,alita2008}, for which a moderate resolving power suffices. The zero noise property has enabled several important experimental milestones in quantum physics\cite{mgiustina2013,mgiustina2015,lshalm2015}. For STJs, which are also pair-breaking detectors, the resolving power has been studied in much more depth, resulting in a measured resolving power of $12-25$ at wavelengths of $1500-400$ nm (best $\delta E$ = 0.07 eV) \cite{dmartin2006}. This corresponds to $R\approx R_{Fano}/2$ in Ta, which has a relatively high $T_c\approx4.4$ K. This resolving power was shown to be partially due to tunnelling noise, which is specific to STJs, and partially due to the loss of hot phonons. These hot phonons are lost during the downconversion process of the initial photon energy to the detected quasiparticles excitations, which involves several stages of electron-electron and electron-phonon interaction. The downconversion process was modelled in detail \cite{mkurakado1981,akozorezov2007,akozorezov2008} and verified experimentally in Ref. \onlinecite{dmartin2006}. To what extent the phonon flow can be impeded during the energy-downconversion and whether that indeed improves the resolving power has not been experimentally addressed.

Here we demonstrate that the resolving power of MKIDs at visible to near-infrared wavelengths is indeed limited by hot phonon losses due to the downconversion process after photon absorption. We use a hybrid NbTiN-Al MKID with aluminum as the sensitive material, which is optimised for this resolving power study. We use a geometry for which we have studied the quasiparticle physics in detail and for which the sensitivity is high enough not to be limited by signal to noise \cite{pdevisser2014,jbaselmans2017,sderooij2021}. We measure the resolving power as a function of photon energy by exposing the detector to continuous illumination of four lasers with wavelengths between $1545-402$ nm. For devices on a SiN/Si substrate and devices on a sapphire substrate, the resolving power is limited to $10-21$ (best  $\delta E$ = 0.08 eV)  for the respective wavelengths, which is consistent with the loss of hot phonons. These numbers are in the same ballpark as the best reported MKID, TES and STJ devices on substrates. When we suspend the sensitive part of the detector on a 110 nm thick SiN membrane, the measured resolving power improves to $19-52$ for $1545-402$ nm (best $\delta E$ = 0.04 eV), due to the strong phonon trapping. The main results are summarised in Fig. \ref{fig:resolvingpower}a. We discuss a route towards reaching the Fano limit by further reducing the phonon flow.

\section{Detector designs}\label{sec:kiddesign}
\begin{figure}
\includegraphics[width=0.99\columnwidth]{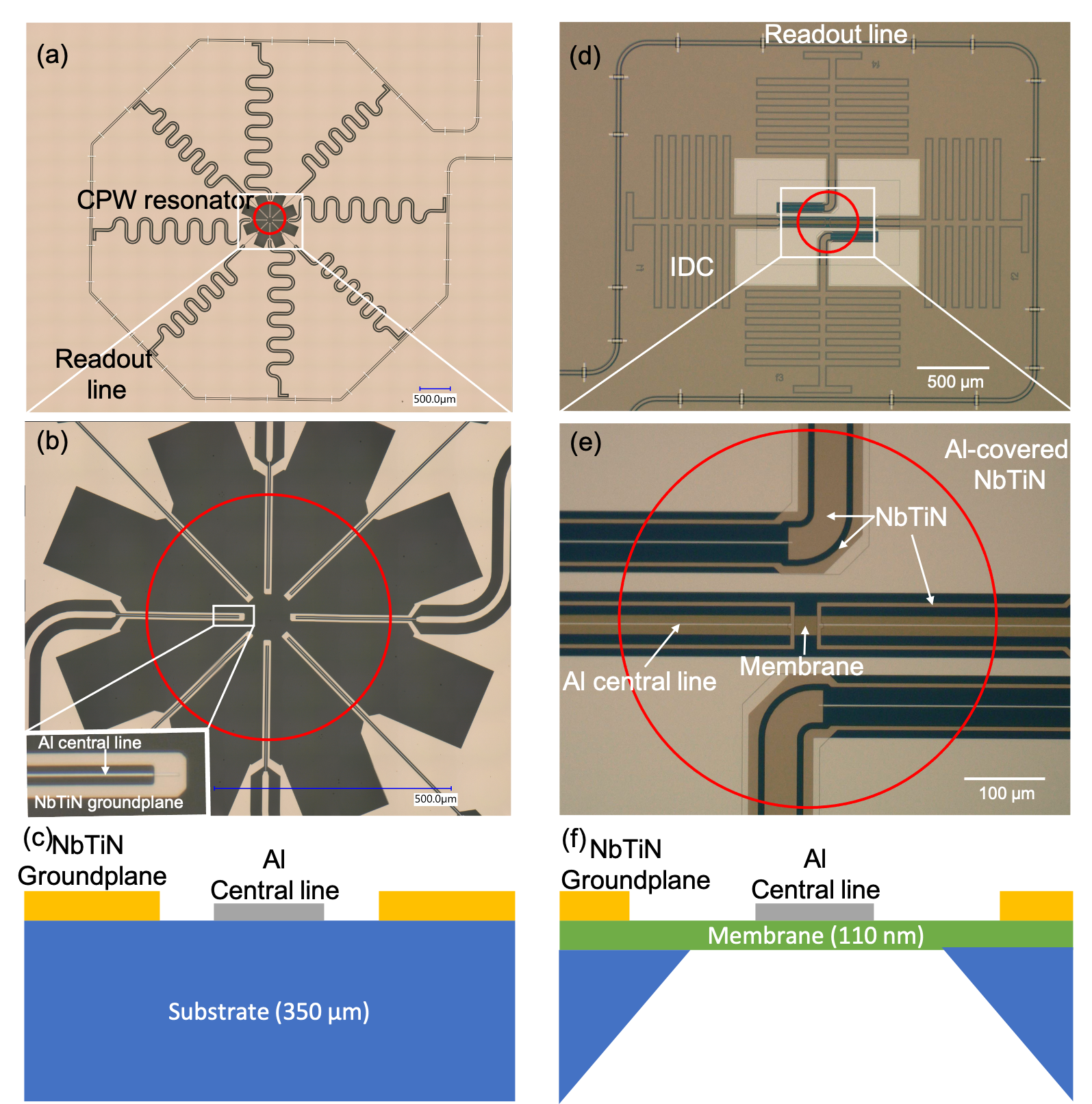}
\caption{\label{fig:designs} (a, b) Micrographs of device A, with 8 CPW resonators which are capacitively coupled to the readout line. The red circle indicates the size and position of the 500 $\mu$m diameter aperture which is mounted in front of the sample and therefore limits the area that is illuminated by laser light. Only near the shorted end of the resonator there is a small section of central line made out of Al (260 or 530 $\mu$m long) which is the sensitive part of the resonator and points towards the centre of the chip. In the centre of the chip most of the NbTiN groundplane is cut away to avoid absorption of light in there. (c) Schematic cross-section of the sensitive shorted end of the detector, a CPW section with NbTiN ground planes and Al central line on a thick sapphire substrate (not to scale). (d, e) Micrographs of devices C on substrate (top and bottom two) and D on membrane (middle two), which have a large interdigitated capacitor (IDC) to enhance the responsivity. The CPW end section is particularly wide to fit the whole membrane width inside. (f) Schematic cross-section of the sensitive shorted end of the detector (not to scale). The 1.7 $\mu$m wide Al strip is fabricated on a 110 nm thick membrane to allow phonons to be re-trapped into the film before escaping to the bath. The NbTiN groundplanes are on the substrate outside the membrane area, to avoid photons absorbed in the NbTiN groundplane to generate a measurable pulse in the Al strip through phonon transport.}
\end{figure}

We use hybrid NbTiN-Al resonators as single photon detectors. The devices used in this work are hybrid NbTiN-Al resonators \cite{jbaselmans2017} and their properties are listed in Table \ref{table:devices}. We use either 350 $\mu$m thick C-plane sapphire substrates (for devices A and B) or 150 nm LPCVD grown SiN on 350 $\mu$m Si (for devices C an D). Sapphire and SiN are isolators with a large enough bandgap to avoid the excitation of charge carriers with 402 nm photons. We use two different designs, one based on the conventional coplanar waveguide (CPW) geometry and a design where a large part of the CPW is replaced by an interdigitated capacitor (IDC)\cite{jbaselmans2021}. Resonators are coupled capacitively to the microwave readout line. Micrographs of the devices are shown in Fig. \ref{fig:designs} (a,b) and (d,e) respectively. In both cases a large part of the resonator is made out of wide NbTiN structures, which strongly reduces two-level system (TLS) noise due to the favourable properties of NbTiN\cite{rbarends2009b} and the width of the structures \cite{jgao2007,jgao2008b}. NbTiN is deposited using reactive magnetron sputtering in an argon-nitrogen plasma \cite{bbos2017,dthoen2017}, with a thickness of 200 nm for devices on SiN and 300 nm for devices on sapphire. Near the shorted end of the quarter wave resonator we put a short section of sputter deposited Al, which is the sensitive part. The ground planes of the readout line are balanced using aluminium bridges on top of a polymide support. The $T_c$ of the NbTiN layer varies somewhat between wafers, and is typically 15 K. The SiN membranes are defined with a KOH etch from the backside of the wafer, after finishing the layers on the front side. After the KOH etch the wafer is cleaned with RCA2. The membrane of device D ends up as 110 nm thick, due to over-etching with SF6 when patterning the NbTiN layer. For Devices C and D the measured Al line width is 1.7 $\mu$m. For phonon trapping in device D we rely on the geometrical effect of phonon re-trapping due to the 1.7/0.110 aspect ratio of the line width and the membrane thickness. We will analyse the resulting improvement in resolving power in Section \ref{sec:phonontrapping}.

For the membrane based device we have found that the number of measured photon hits is strongly dominated by hits in the NbTiN groundplane strip, when that NbTiN strip is also placed on the membrane. Therefore, we have made the end-section of these resonators much wider, to only have the central Al strip on the membrane (Fig. \ref{fig:designs}f). The reference detector C on the substrate has the same geometry to allow a direct quantitative comparison. For these wide sections we have found that the signal-to-noise is degraded about 3x compared to a narrow CPW (Fig. \ref{fig:designs}b), which is compensated by the use of the IDC in these designs. As will be shown in Section \ref{sec:eresolution}, the designs indeed have sufficient signal-to-noise to study the intrinsic limits to the resolving power. 

We have chosen these designs to study the resolving power and the effect of phonon losses in a device that we understand well. The downside of this geometry, together with the choice for Al as active material, is that the absorption efficiency of photons (or quantum efficiency) of the devices is low. The efficiency is limited by the Al fill fraction in the aperture and the Al surface impedance at the measured wavelengths. 

\begin{table*}[t]
\centering
\begin{tabular}{ccccccc}
\hline
Device	 & Substrate			& Al thickness 	& Al width 		& Resonator type 	& $T_c$ (K) & $\rho$ ($\mu\Omega$cm) 	\\ 

\hline
$A$		&  Sapphire			& 50 nm		& 0.9 $\mu$m	& CPW	& 1.25*	& 0.9*	\\ 	
$B$		&  Sapphire			& 150 nm	 	& 0.7	 $\mu$m	& CPW	& 1.12	& 0.4		\\ 	
$C$		&  SiN on Si			& 50 nm		& 1.7 $\mu$m	& IDC	& 1.25 	& 1 		\\   	
$D$		&  110 nm SiN membrane		&  50 nm		& 1.7 $\mu$m	& IDC	& 1.25 	& 1 	\\	
\hline
\end{tabular}

\caption{Overview of the measured devices. Devices A and B have 8 resonators on a chip. Devices C and D are fabricated on the same chip. We explicitly verified that Al $T_c$ and $\rho$ are the same on SiN membrane and substrate. The Al linewidths are measured from scanning electron microscope (SEM) images of each chip. For device A, the co-fabricated DC structure had a design error. The data shown here was measured on a device with the same geometry, Al thickness, substrate and fabrication steps.}\label{table:devices}
\end{table*}

The design excludes two possible limitations to the energy resolution. Quasiparticle diffusion out of the sensitive part of the detector could cause pulse height variations. $\Delta_{\textrm{NbTiN}}/\Delta_{\textrm{Al}}\approx 12$, which prevents the quasiparticles generated in the Al to diffuse out of the Al volume during the single photon pulse. Secondly microwave current density non-uniformity in the detector could lead to a different pulse height for the same photon energy when absorbed in different locations \cite{bmazin2012,nzobrist2019}. The current density in the Al part of our resonators is uniform to 99\%. Together with quasiparticle diffusion \textit{inside} the Al volume, which leads to a uniform quasiparticle density, we expect current non-uniformity not to play a role for $R<100$.

To limit absorption of radiation in unwanted areas on the chip we place a 500 $\mu$m diameter aperture on top of the chips, as indicated with a red circle in Figure \ref{fig:designs}. Since NbTiN absorbs a lot more visible/near-infrared light than Al, which is unwanted, we cut away most of the NbTiN groundplane inside the aperture for devices A and B. For devices C and D we cover most of the groundplane with Al, since absorption in the Si substrate (with a 1.2 eV bandgap) is also undesirable.

\section{Experimental setup}\label{sec:setup}
\begin{figure}
\includegraphics[width=0.9\columnwidth]{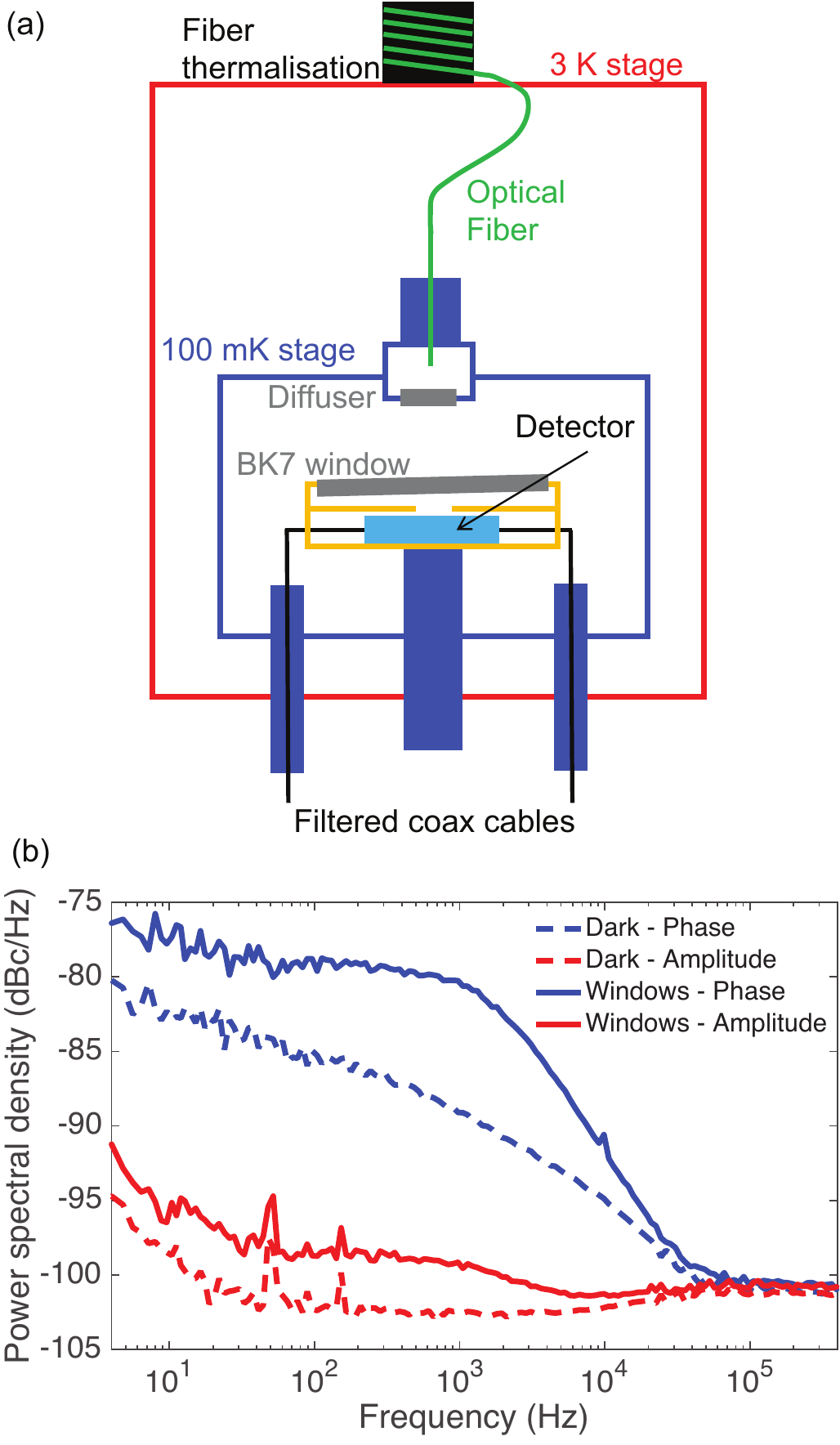}
\caption{\label{fig:setup} (a) Schematic of the inner parts of the cryostat, showing the box-in-box concept where the only hole at 100 mK is made for the optical fiber. A detailed description can be found in the text. (b) Noise spectra of the phase and amplitude response of a CPW MKID in a dark environment and in an environment with open (BK7 and fused silica) windows to the 3 K stage of the setup. Both measurements are performed without laser illumination.  }
\end{figure}

The samples are cooled in a pulse-tubed pre-cooled adiabatic demagnetisation refrigerator. Measurements are carried out at a temperature of 120 mK, unless otherwise indicated. The sample stage is carefully shielded from stray light from the 3 K stage of the cooler, using a box-in-box concept with coax cable filters in the outer box\cite{jbaselmans2012,pdevisser2014}. The system is schematically depicted in Figure \ref{fig:setup}(a). Laser light is coupled into the the cryostat with an optical fiber (Thorlabs SMF-28-J9, with a 8.2 $\mu$m diameter core and 125 $\mu$m cladding, which is single-moded at 1550 nm). The fiber is thermalised at 3 K with 8 windings on a copper spool with a 32 mm diameter. The fiber ends at the 100 mK box as a bare cleaved end, which is mounted with a custom copper clamp, to avoid standard commercial parts inside the magnetic shields, which are typically made from stainless steel (i.e. magnetic). The only possible entry for (stray) light into the 100 mK box is therefore through the fiber. Light enters the light tight box through a fused silica engineered diffuser (RPC Photonics), which spreads out the narrow beam from the fiber over an angle of 32 degrees, and also compensates for multimode patterns which can occur for the shorter wavelengths. At the sample box we mount a 5 mm thick BK7 glass window. Right above the sample we place a 500 $\mu$m diameter aperture as described above. Outside the aperture light is absorbed using black, carbon loaded epoxy. The samples are front-illuminated.

Microwave measurements are performed with a standard homodyne detection scheme. The signal from the microwave generator is first attenuated at every temperature stage, sent through the sample and amplified with a low noise HEMT amplifier at 3 K and with a room temperature amplifier. The noise floor we observe in the measurement corresponds to an effective system noise temperature of about 5 K. The output is mixed with the original signal using an IQ mixer. A detailed overview of all components can be found in Ref. \onlinecite{pdevisserphd}.

In an early experiment of this kind, the optical fiber was not mounted at the lowest temperature stage, but ended at the 3 K stage of the cooler. The detector, illuminated through the same fused silica diffuser and BK7 glass window (both at 120 mK), was therefore also exposed to the 3 K environment. In this configuration we found a strong increase in the noise, compared to a dark measurement with the sample box closed with a copper lid, as shown in Figure \ref{fig:setup}(b). This demonstrates that the thermal radiation of the 3 K environment is not filtered by the fused silica and BK7 glass windows and is visible as photon noise in the detector response. Although BK7 is one of the best possible glass filters for wavelengths from near-infrared to sub-millimeter, it is known to be transparent for frequencies lower than $\sim$500 GHz \cite{mnaftaly2007}. The Al detector is sensitive to any light with frequencies higher than its gap frequency of 92 GHz. For the experiments reported here we have solved this by mounting the optical fiber directly to the 120 mK stage as described above. For imaging experiments with these detectors through a window, a low-frequency filter will be indispensable to preserve the resolving power. 

\section{Measurement procedure}\label{sec:measproc}
For every new measurement condition, we first perform an $S_{21}$ frequency sweep to find the resonant frequency and to measure the resonance circle as a reference for the MKID response. At the resonant frequency we first perform a noise measurement with the lasers off at 1 Msample/s. For the resolving power measurements, we use four room temperature, fiber-coupled diode lasers (Thorlabs) with wavelengths 1545, 986, 673 and 402 nm, with output powers between $1-10$ mW in continuous wave mode and a linewidth corresponding to $R>300$. To enable single photon absorption at a rate of $\sim$100 Hz, around 80 dB of attenuation is used before entering the cryostat. Note that also laser power is lost due to the inefficient absorption as explained above and because the illumination spot after the diffuser is larger than the aperture. We take 40 s of data for devices A and B and 160 s for devices C and D, sampled at 1 Msample/s, which aims at a few thousand single photon events per wavelength. 

\begin{figure*}
\includegraphics[width=0.9\textwidth]{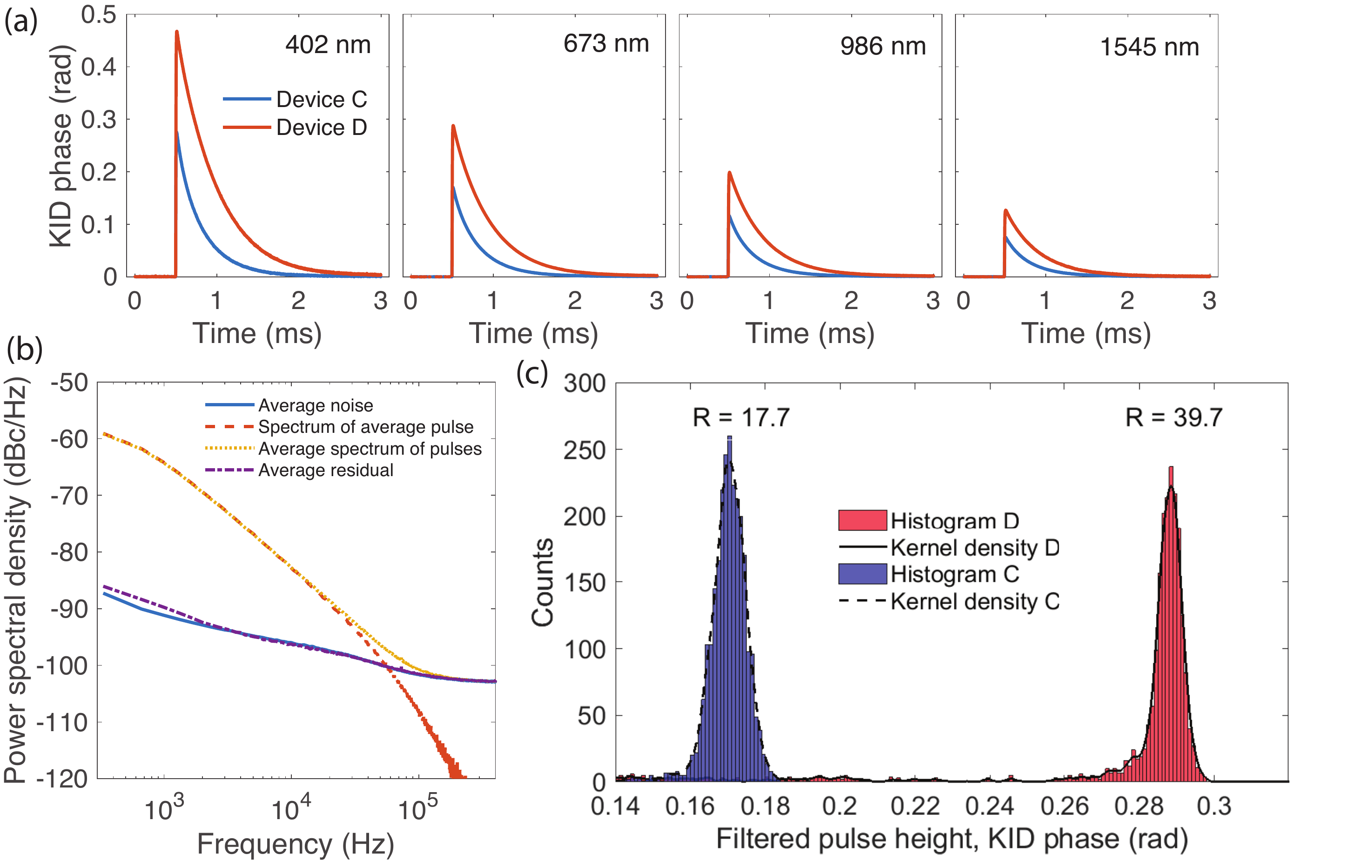}
\caption{\label{fig:optimalfilter} (a) Average pulse shape, obtained by averaging typically thousand single photon pulses measured at the indicated wavelengths for devices C (substrate) and D (membrane). The pulse height is partially due to the different $Q_c$ (37.6$\times 10^3$ for C and 46.0$\times 10^3$ for D), but mainly due to the stronger phonon trapping in device D. (b) Power spectral densities of the average noise (solid) and average pulse (dashed), which together constitute the optimal filter for device C at 120 mK, with $P_{read}$ = -83 dBm illuminated with a 673 nm laser. The area in between the pulse and noise curves determines the effective $R_{SN}$ and bandwidth. As a consistency check we also plot the average of each pulses spectrum (dotted) and average residual, $\theta(f) - AM(f)$ (dash-dotted). (c) Histogram of pulses of device C and D at 120 mK, illuminated with the 673 nm laser and with $P_{read}$ = -83 dBm and $P_{read}$ = -85 dBm respectively. The comparison of the FWHM of these histograms demonstrates the strong improvement in $R$ due to phonon trapping from $R = 17.7$ to $R = 39.7$. The change in the average pulse height in the histogram corresponds to that in panel (a). }
\end{figure*}

Single photon pulses are processed with a standard optimal filter, applied to the MKID phase response, $\theta$, which gives the highest signal-to-noise for our devices. The optimal filter is applied in the frequency domain, to extract the optimal pulse amplitude $A$ from the measurement \cite{kirwinphd}. We consider $\theta(f) = AM(f)+N(f)$, with $\theta(f)$ the Fourier transform of the measured pulse, $M(f)$ the pulse model, normalised to 1, and $N(f)$ the noise. We can estimate $A$ using \cite{kirwinphd}
\begin{equation}
A = \int_{-\infty}^{\infty} \frac{\theta(f)M^*(f)}{|N(f)|^2} df \bigg/ \int_{-\infty}^{\infty} \frac{|M(f)|^2}{|N(f)|^2} df,
\label{eq:optimalfilter}
\end{equation}
where the bandwidth is limited at low frequencies by the duration of the pulses, which we take 3 ms, including 0.5 ms of pre-pulse data, and at high frequencies by the sampling rate. Time traces that have multiple pulses (on top of each other) within this time window are left out of the analysis. The pulse component at $\theta(f=0)$ is left out of the analysis since it does not improve, or even deteriorates the resolving power. The pulse template $M(f)$ is calculated for each measurement condition as the average of the detected pulses, where we iterate once to only use pulses which have a pulse amplitude within 3$\sigma$ around the maximum in the histogram. Typical average pulses are shown in Fig. \ref{fig:optimalfilter}a for all wavelengths for devices C and D. The noise power spectral density $|N(f)|^2$ is calculated from the time stream with lasers off. We average the spectrum of thousand samples with the same length as the pulses. Care is taken to filter out large energy events that also occur with lasers off, due to e.g. cosmic rays \cite{pdevisser2011,kkaratsu2019}, before calculating $|N(f)|^2$. The average pulse and noise spectra are shown for device C in Fig. \ref{fig:optimalfilter}b. In the plot we distinguish between the spectrum of the average pulse (averaging the time domain pulses before taking the spectrum) and the average spectrum of pulses (averaging the spectra of each pulse) to check that the noise floor in the pulse measurements is the same as in the noise measurement. The 'average residual' represents a spectrum of the time traces where pulses were detected, after subtracting the optimally filtered pulse, $\theta(f) - AM(f)$, to check the consistency of the analysis.

In the analysis we distinguish several aspects of the resolving power, expressed by different symbols. $R$ is the measured resolving power of the detector. A typical histogram for 673 nm illumination of devices C and D is shown in Fig. \ref{fig:optimalfilter}c, which already illustrates the large difference in $R$ for substrate and membrane based devices, to which we will come back in Section \ref{sec:eresolution}. $R$ is calculated as the FWHM of a kernel density estimate, shown as the lines in Fig. \ref{fig:optimalfilter}c, where we choose the kernel bandwidth to be the same as the bin-size of the histogram. In Appendix \ref{app:lowenergyhits}, we show histograms for all wavelengths and for the whole energy range. In particular for the membrane device we also observe low energy hits due to the non-selective illumination.

$R_{SN}$ is the estimate of the signal-to-noise contribution to the resolving power from the noise spectrum and the pulse template\cite{kirwinphd,meckartphd}: 
\begin{equation}
R_{SN} = \frac{\overline{A}}{2\sqrt{2\ln2}}\sqrt{\int_{-\infty}^{\infty} \frac{|M(f)|^2}{|N(f)|^2} df},
\end{equation}
with $\overline{A}$ the average pulse height. $R_{SN}$ is also called $R_{NEP}$ in literature. $R_0$ is the measured resolving power using the optimal filter (Eq. \ref{eq:optimalfilter}) on pieces of noise. The resulting histogram is centred around $E=0$ and $R_0$ represents the FWHM with respect to the photon energy, $R_0 = E/\delta E_{E=0}$. It is a measure of the potential influence of e.g. baseline fluctuations that are not taken into $R_{SN}$. Finally we will analyse the contributions to the measured $R$ using $1/R_{intrinsic}^{2} = 1/R^{2}-1/R_{SN}^{2}$. $R_{intrinsic}$ is a measure of the resolving power contributions that are not due to noise.

\section{Pulse height analysis}\label{sec:pulses}
Figure \ref{fig:optimalfilter}a shows the pulse response to single photons at the four measured wavelengths, each averaged over a few thousand single photon hits, for device C on SiN substrate and device D on membrane. 

We can estimate the pulse height using the number of quasiparticles generated by the photon $N_{qp} = \eta_{pb}E/\Delta$. For a linear detector, the resulting detector phase response is then given by $\theta = N_{qp} \frac{d\theta}{dN_{qp}}$. We calculate the responsivity factor using $\frac{d\theta}{dN_{qp}} = -\frac{4Q}{f_{res,0}}\frac{df_{res}(T)}{dN_{qp}(T)}$, where the resonant frequency $f_{res}(T)$ is obtained from a measurement of $S_{21}$ as a function of bath temperature \cite{jbaselmans2008}. $Q$ is taken from the $S_{21}$ measurement that we perform before each pulse measurement. This leaves $\eta_{pb}$ as the only unknown, which we therefore use as a fit parameter. We calculate $\eta_{pb}$ for each wavelength and for three different $P_{read}$, which results in an average $\eta_{pb} = 0.30\pm0.02$ for device C and $\eta_{pb} = 0.40\pm0.04$ for device D. The uncertainty represents the standard deviation of the twelve measurements for each device. For both devices, lower $P_{read}$ leads to higher pulses and therefore higher estimated $\eta_{pb}$. The difference in $\eta_{pb}$ between devices C and D means that in device D $\sim$33\% more energy is kept in the film during the downconversion process. The numbers are close to the expected $\eta_{pb}$ \cite{akozorezov2000,tguruswamy2014}, although it has not been calculated in aluminum at high energies. For perfect phonon trapping, $\eta_{pb} = 0.59$ is expected. For the trapping factor on membrane of $\sim$5 we have here, this maximum is not yet reached, as shown in Ref. \onlinecite{tguruswamy2014} for low energies. The present experiment should not be taken as a measurement of $\eta_{pb}$ (as in Ref. \onlinecite{pdevisser2015}), because of the large range of energies involved, but mainly as a verification that the single photon pulse heights in Al MKIDs are consistent with our current understanding of the response.

Capturing the behaviour of the resonator in a single responsivity term $d\theta/dN_{qp}$, measured for a quasiparticle population at elevated bath temperature is a strong assumption. A more detailed pulse model, including the pulse rise and decay, should start from the rate equations for quasiparticle recombination \cite{arothwarf1967,cwilson2004}, with an excitation due to the absorbed photon energy. The resulting $N_{qp}(t)$ can be converted to complex conductivity \cite{dmattis1958}, $\sigma_1(t)$ and $\sigma_2(t)$, and surface impedance. Together with an estimate of the kinetic inductance fraction $\alpha_k$ \cite{jgao2008c}, and the measured quality factors, one can calculate $S_{21}(t)$. The only leftover fit parameter would also be $\eta_{pb}$. We have found however that the results of such a model strongly depend on whether one interprets the observed saturation in the pulse decay times at low temperatures as an effective quasiparticle temperature, or as a saturation of the lifetime-only. This lifetime saturation is commonly observed \cite{rbarends2009,pdevisser2011,jzmuidzinas2012}. We have recently performed new experiments, part of which on the here presented devices, which support a lifetime saturation without quasiparticle number saturation \cite{sderooij2021, sderooijmsc}. 

\section{Energy resolution on substrate and on membrane}\label{sec:eresolution}
\begin{figure*}
\includegraphics[width=0.99\textwidth]{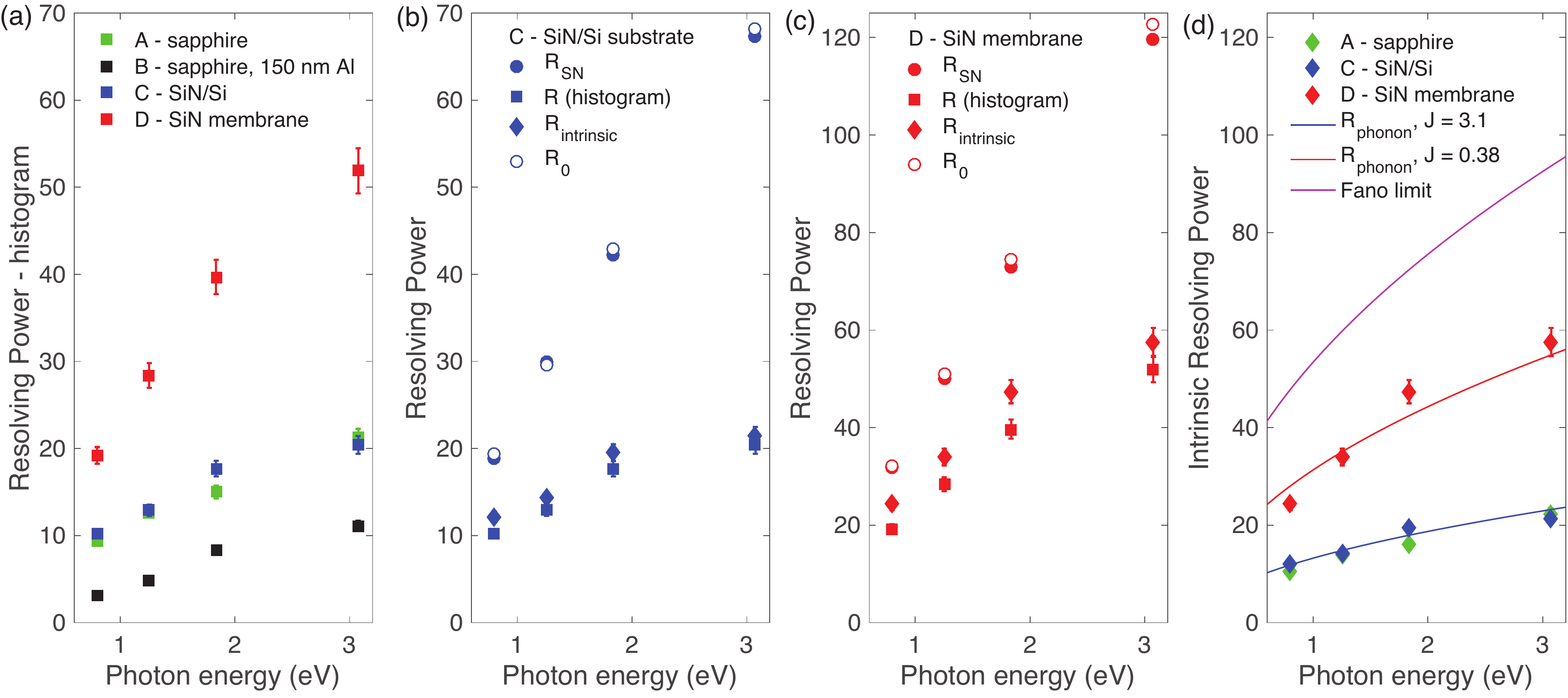}
\caption{\label{fig:resolvingpower} Measured resolving power. (a) The resolving power $R$ as obtained from the histogram of pulses for each of the four measured wavelengths for the devices A-D. The much higher $R$ due to phonon trapping for the device on a membrane is the main result of this work. The error bars are described in the text. (b,c) The different contributions to the resolving power for devices C and D respectively. The $R_{SN}$ is the signal-to-noise contribution obtained from the average pulse and noise (filled circles). $R_0$ represents the FWHM of the histogram when the optimal filter is applied on traces of noise (open circles). $R$ is the measured histogram resolving power as shown also in panel (a) (filled squares). $R_{intrinsic}$ quantifies all contributions to $R$ that are not due to signal-to-noise, in this case dominated by hot phonon losses. (d) $R_{intrinsic}$ for devices A, C and D. The $R_{phonon}$ lines are fits of Eq. \ref{eq:phonon} to the data and represent the expected $R$ for different levels of hot phonon loss. The magenta line is the Fano limit for the Al film properties of device D in Table \ref{table:devices}, using Eq. \ref{eq:Fano}. Note that the y-scale of the panels is different. }.
\end{figure*}

The measured resolving power for the different devices as a function of photon energy is shown in Fig. \ref{fig:resolvingpower}a, which is the main result of this work, together with the characteristic histogram in Fig. \ref{fig:optimalfilter}c. We observe that for devices A and C on sapphire and SiN substrates the resolving power is similar and limited to $10-21$ for wavelengths $1545-402$ nm, respectively. We also observe that roughly $R\propto\sqrt{E}$, which already hints upon an intrinsic effect, since $R_{SN}\propto E$. For device D on membrane, the resolving power increases to $19-52$, which is a large improvement compared to devices A and C, and demonstrates that $R$ on substrate is indeed limited by the loss of hot phonons. The measured $R$ at 1545 nm corresponds to a best measured energy resolution of 0.04 eV. The error bars are obtained from varying the kernel width of the kernel density estimate from a too coarse to a too fine representation of the histogram, which can lead to a $\pm$5\% variation in the estimated $R$.

On the way to a higher $R$, we have also measured a 150 nm Al film on sapphire, device B. The phonon escape time increases linearly with film thickness, which should enhance a phonon-loss limited $R_{intrinsic}$ significantly. However, as we observe in Fig. \ref{fig:resolvingpower}a, the measured $R$ is even lower than for devices A and C. $R_{SN}$ strongly limits $R$ for this thick film due to the reduced responsivity, which is caused by the increase in Al volume and decrease in $\alpha_k$.

We separate the contributions of $R_{SN}$ and $R_{intrinsic}$ to $R$ as described in Section \ref{sec:measproc}, which is shown in Fig. \ref{fig:resolvingpower}b for device C and in \ref{fig:resolvingpower}c for device D (the y-scale of panels a,b is different from panels c,d). We observe that going from substrate to membrane increases $R_{SN}$ by a factor $1.7$, which is due to the increased pulse height on the membrane, where the noise stays the same (see Fig. \ref{fig:spectraGR} and Section \ref{sec:discussion}). We also observe that $R_{SN}$ is close to linear in $E$, as expected in the linear regime of the MKID response, and as expected from the pulse heights in Fig. \ref{fig:optimalfilter}a. 

In both devices C and D, $R_{SN}$ is a small contribution to $R$ and so $R_{intrinsic}$ dominates. We have excluded known limits to $R_{intrinsic}$ due to quasiparticle out-diffusion and nonuniform current density by design (Section \ref{sec:kiddesign}). It is also unlikely that baseline fluctuations that are not captured into $N(f)$ play a dominant role since $R_0$ is very close to $R_{SN}$ for both devices. It is therefore plausible that $R_{intrinsic}$ for the membrane device is also dominated by hot phonon losses. We therefore interpret the difference between substrate and membrane as entirely due to the enhanced phonon trapping.

\section{Analysis of phonon loss and phonon trapping}\label{sec:phonontrapping}
We first derive the experimental phonon loss factors from the resolving power measurements in Section \ref{subsec:J} and analyse the relative phonon trapping between devices on substrate and on membrane in Section \ref{subsec:Jrelative}. In Section \ref{subsec:Jabsolute} we analyse the absolute value of the phonon loss factor for the substrate devices.
\subsection{Measured phonon loss factor}\label{subsec:J}
We quantify the effective level of phonon loss and phonon trapping on the membrane by comparing the $R_{intrinsic}$ for the devices on substrate (C) and on membrane (D), as shown in Figure \ref{fig:resolvingpower}d. The influence of hot phonon loss on the resolving power can be expressed as \cite{akozorezov2007,akozorezov2008}
\begin{equation}
R_{phonon} = \frac{1}{2\sqrt{2 \ln2}}\sqrt{\frac{\eta_{pb}^{max}E}{\Delta(F+J)}},
\label{eq:phonon}
\end{equation}
with $J$ accounting for the hot phonon loss. Per definition the Fano limit is reached when $J=0$ (Eq. \ref{eq:Fano}). Note that the minimum energy to generate a quasiparticle, $\Delta/\eta_{pb}^{max}$, should be used here also for significant phonon losses. The generally used $\eta_{pb}$ is an expression of average energy lost. $R_{phonon}$ expresses the variance of that process, for which the effect of phonon loss is taken into account by $J>0$. 

Using the known values for $F$, $\Delta$, and $\eta_{pb}^{max}$, and assuming that $R_{intrinsic}$ is indeed dominated by phonon losses, we can fit Eq. \ref{eq:phonon} to $R_{intrinsic}$ for devices C and D, which results in the curves plotted in Fig. \ref{fig:resolvingpower}d, with $J_C = 3.1\pm0.4 $ and $J_D = 0.38\pm0.08 $. We conclude that in device D, hot phonon loss still dominates over the Fano factor ($F=0.2$).

\subsection{Enhanced phonon trapping on membrane}\label{subsec:Jrelative}
The phonon loss factor $J$ is expected to scale linearly with the phonon trapping factor $\tau_{esc}/\tau_{pb}$, with $\tau_{esc}$ the average time a phonon spends in the superconductor and $\tau_{pb}$ the average time it takes to break a Cooper pair \cite{akozorezov2007,akozorezov2008}. $\tau_{pb} = 0.26$ ns for Al \cite{skaplan1976}. The ratio $J_C/J_D = 8\pm2$ derived from the experiments therefore expresses the \textit{relative} phonon trapping enhancement between the device on membrane compared to the substrate. 

To model the enhanced phonon trapping we consider that the phonon energies which dominate the downconversion process range from the Debye energy to about $10\Delta$ \cite{akozorezov2000}. These energies correspond to phonon wavelengths of $0.5-10$ nm, which are all smaller than the dimensions of our device. Therefore we use a geometrical ray tracing model, in which phonons are generated in the superconducting film with random angles. Upon each interaction with the interface we calculate the transmission probability from the acoustic matching\cite{skaplan1979}. A phonon which ends up in the substrate is lost. For the membrane a phonon has a (high) chance to be reflected back into the film, dependent on the membrane thickness and the width of the film. After the phonon leaves the membrane area underneath the superconductor, we introduce a 50\% chance of return to account for diffusive reflection at the membrane surfaces. Details of the model are discussed in Appendix \ref{app:phonontrapping}. The resulting average phonon dwell time in the 1.7 $\mu$m wide Al strip is 0.16 ns for the substrate and 1.3 ns for a 110 nm membrane. The resulting phonon trapping enhancement is a factor 8.1, which agrees well with the value derived above from the resolving power measurement, despite the simplicity of the model.

\subsection{Absolute phonon loss calculation on substrate}\label{subsec:Jabsolute}
So far we only compared the relative $R_{intrinsic}$ between substrate and membrane. The absolute value of $J_C$, on substrate, can be compared to the detailed model of the different downconversion stages in Refs. \onlinecite{akozorezov2000,akozorezov2007,akozorezov2008}. 

For the high energy part of the downconversion process (from $E_{photon}$ to around $E_{Debye}$), we evaluate Eq. 25 from Ref. \onlinecite{akozorezov2008}, for photons incident on the top-surface. We first consider that for the highest phonon energies, around the Debye energy, only longitudinal phonons are available for Al \cite{ssavrasov1996,akozorezov_comm}, and therefore the input parameters are taken only for longitudinal phonons. The result is $J_{high} = 2.8$. The input parameters were taken for Al from Table \ref{table:devices} and the timescales and characteristic energies from Refs. \onlinecite{akozorezov2000,skaplan1979} and the acoustical properties of SiN from Ref. \onlinecite{tkuhn2004}. 

The low energy part of the downconversion involves energies ranging from $E_{Debye}$ to the characteristic energy for Al, which is 10.6$\Delta$ \cite{akozorezov2000}. We use Eq. 6 from Ref. \onlinecite{akozorezov2008}, with a correction \cite{akozorezov_comm} as given in Appendix \ref{app:eqdownconversion}. In this phase both longitudinal and transversal phonons play a role. The result from this calculation is $J_{low} = 0.24$. It is clear that the dominant effect on the energy variance, and therefore on the resolving power, originates from the first steps of the downconversion process as expressed in $J_{high}$.

The phonon loss factors from both stages can be added together, $J=J_{high}+J_{low} = 3.0$, which agrees well with the measured $J_C$. 

When we fit in the same way the measured $R_{intrinsic}$ for Device A on sapphire, we obtain $J_A = 3.7\pm0.3$. We then use the same equations with the acoustical properties of sapphire to obtain theoretically $J=2.7$, which is lower than the measured $J_A$.

We realise that not all the input parameters to calculate $J_{high}$ and $J_{low}$ are accurately measured. For example, we have assumed a diffusion constant for thin Al of 20 cm$^2$/s, where measured values range from 18-60 cm$^2$/s \cite{cwang2014,shsieh1968,sfriedrich1997}. Next to that, the Debye model is poorly justified for the high energies involved. Additionally, the acoustic model assumes a perfect interface for transmission and reflection, which may be rough in reality. Therefore a realistic error margin on the calculation for $J$ is at least $\pm$0.5. In the light of these considerations and the large range of energies involved, a correct order of magnitude estimate is already satisfactory.

\section{Discussion}\label{sec:discussion}

\begin{figure}
\includegraphics[width=0.99\columnwidth]{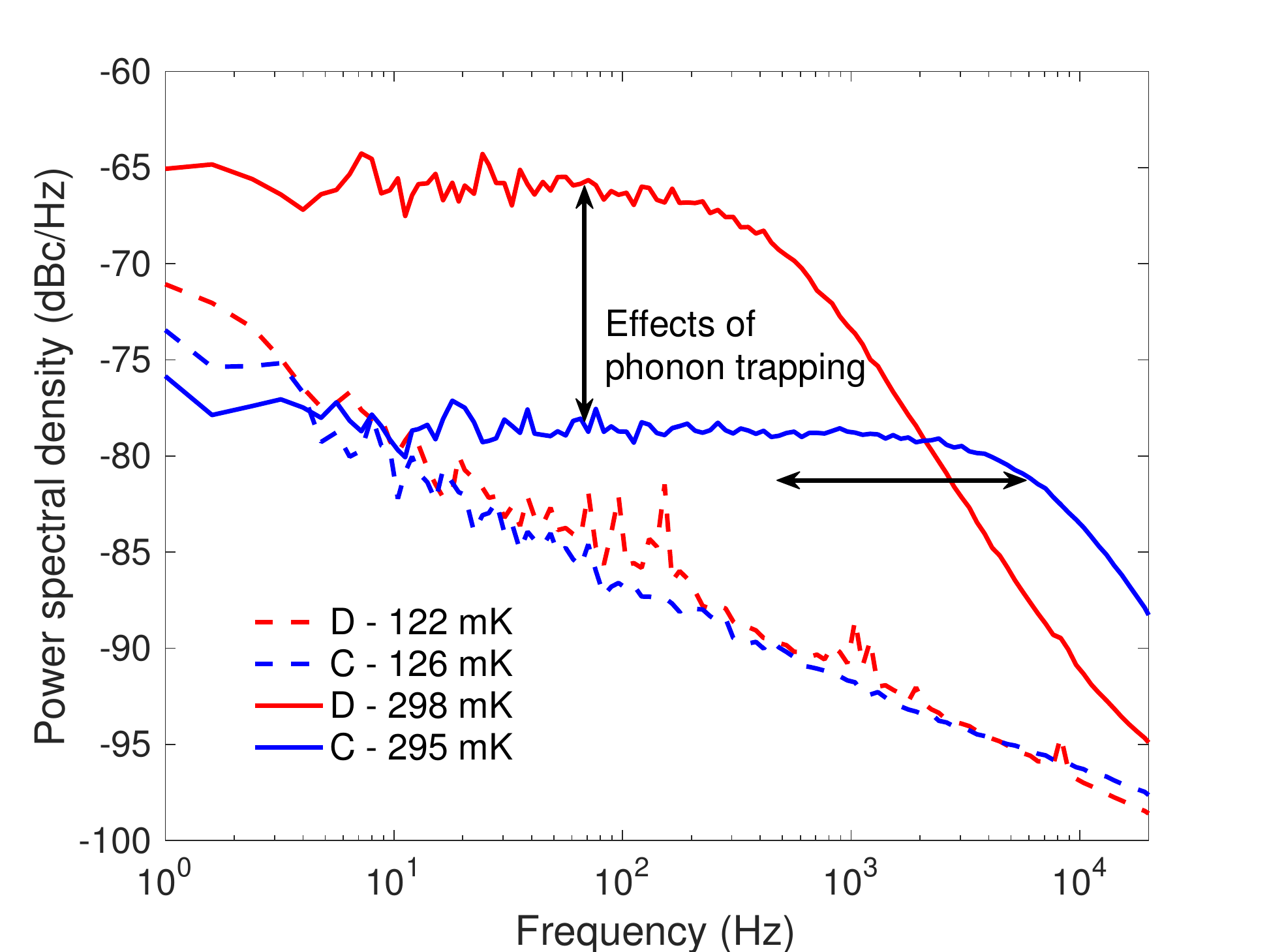}
\caption{\label{fig:spectraGR} Power spectral density of the resonator phase for devices C and D measured at temperatures close to 120 mK (dashed lines), where we measured the resolving power, and at high temperatures (solid lines). The arrows indicate the enhanced generation-recombination (GR) noise level and lifetime due to phonon trapping, which show up at high temperature. The low temperature spectra have no signature of GR noise.}
\end{figure}

\subsection{Phonon trapping factor in quasiparticle fluctuations}
We have also measured quasiparticle fluctuations (generation recombination noise) in the same devices, which we present in detail in Ref. \onlinecite{sderooij2021}. The quasiparticle recombination times on substrate and membrane, fitted from the spectra of these fluctuations, show a factor $66\pm17$ difference in phonon escape time. The effect is illustrated for the high temperature ($T\sim300$ mK) noise spectra of devices C and D in Fig. \ref{fig:spectraGR}. The phonon trapping factor that applies to quasiparticle fluctuations and lifetime in the spectra is $(1+\tau_{esc}/\tau_{pb})$. The quasiparticle lifetime is a factor 16 longer and the noise 12 dB higher on the membrane. The discrepancy between the numbers for $\tau_{esc}$ from the measured $R$ and from the quasiparticle fluctuations needs further study. The quasiparticle fluctuations are maintained by phonons with energies close to $2\Delta$, much lower than those that determine $R$. At the corresponding phonon wavelength, $\sim$50 nm, the phonon dimensionality and the distribution over different phonon modes starts to play a role, for which the ray tracing model is inappropriate. 

As shown in Fig. \ref{fig:spectraGR}, the effect of phonon trapping is clearly visible in the GR noise spectra at higher temperatures ($>$250 mK). However, we observe roughly the same lifetime for the main part of the pulse decay in both cases at 120 mK (319 $\mu$s for C and 431 $\mu$s for D). At the same time, we also observe a strong reduction in generation-recombination noise at low temperatures, as shown for temperatures close to 120 mK in Fig. \ref{fig:spectraGR}, which was unexpected \cite{cwilson2004,pdevisser2011}. This is nevertheless the main reason why we get the required $R_{SN}$ for a membrane-based device at 120 mK. Both these observations and their possible relation with quasiparticle trapping are extensively discussed in a separate publication \cite{sderooij2021}. For comparison we have measured the resolving power for device D at 250 mK at 1545 nm. At this temperature $R = 11$, limited by $R_{SN}=13.8$, which is much lower than at 120 mK, despite a longer pulse decay time of 1.0 ms ($R_{SN}=31.8$ at 120 mK). Understanding this reduction in noise is important to understand the high $R_{SN}$ measured here, but we emphasise it does not influence the limit to $R$ due to hot phonon loss.

\subsection{Towards a Fano-limited resolving power}
The most important next question that is raised by this work is how to experimentally reach a Fano limited energy resolution. In the case of Al on a thin membrane as in device D, we have $J_D=1.9F$, whereas $J\ll F$ is needed for the Fano factor to dominate the resolving power. If we set a goal of $R=0.9 R_{Fano}$, we need $J<F/5$, and therefore another factor of 10 better phonon trapping. This is unrealistic with the geometrical phonon trapping approach we take now, since either the membrane becomes impractically thin, or the Al strip becomes too wide to conserve $R_{SN}$ (only with a more resistive metal, a wider strip remains an option). We therefore envision the fabrication of a phononic crystal (PnC) to better trap the phonons that flow to the bath. It is important for MKIDs to have a PnC that blocks high frequency phonons, but is transparent for phonons at the microwave readout frequency ($\sim2-6$ GHz) to reduce microwave non-equilibrium effects \cite{pdevisser2014b,lcardani2018}. It is possible to achieve a factor 10 phonon trapping at particular energy bands (compared to an unpatterned membrane), while keeping high phonon transparency at low energies \cite{krostem2018,tpuurtinen2020}. 

The analysis above shows that it is realistic to get much closer to the Fano limit. However, reaching $R>0.9R_{Fano}$ is unrealistic from a phonon-trapping point of view. Additionally the requirement $R_{SN}\gg R_{Fano}$ is also difficult. For our Al MKIDs, apart from the additional phonon trapping, we aim to improve $R_{SN}$ further by reducing the film thickness to lower the volume and improve the kinetic inductance fraction. Additionally we can use a pulse filter combining the amplitude and phase response \cite{nzobrist2019,lcardani2017}. For the present data, we found the $R_{SN}$ of phase-only to be sufficient.


It is striking that most substrate-based MKIDs with a high enough $R_{SN}$ are all limited to $R\approx10-20$ \cite{bmazin2012,bmazin2013,pszypryt2017,nzobrist2019,nzobrist2019b,nzobrist2021}. Here we conclusively show that in Al MKIDs on substrate $R$ is limited by hot phonon loss. For other materials this is likely the case, although the acoustic properties of these materials need to be better characterised. We emphasise that the loss of hot phonons arises at the initial stages of downconversion at energies much higher than the quasiparticle excitations that are eventually detected. Therefore, hot phonon loss arises in pair-breaking detectors (MKID, STJ, QCD), in thermal (e.g. TES) detectors, and also in SNSPD detectors\cite{akozorezov2017}. 

\section{Conclusion}\label{sec:conclusion}
We have measured the resolving power of superconducting, energy-resolving, single photon counting MKIDs for wavelengths of $1545-402$ nm. By comparing devices  where the superconductor is deposited on substrate with devices on a 110 nm membrane, we show that the resolving power is limited by the loss of hot phonons. These phonons are involved in the downconversion process, in which the absorbed single photon energy is converted into measurable quasiparticle excitations through multiple stages of electron-electron and electron-phonon interaction. For devices on substrate the resolving power is limited to $10-21$ for $1545-402$ nm respectively. On the membrane this improves to $19-52$, which is consistent with a factor $8\pm2$ stronger phonon trapping compared to the substrate. Even stronger phonon trapping, possibly using phononic crystals, is needed to get closer to a Fano-limited resolving power, together with a further improvement of the signal-to-noise (thinner film, smaller volume).


\section*{Acknowledgements}
We acknowledge help with the experimental setup and initial data analysis from Max van Strien, Robert Huiting, Klaas Keizer and Juan Bueno. We acknowledge help from Alex Kozorezov to interpret his equations for the downconversion process, and insightful discussions with Karwan Rostem, Tuomas Puurtinen, Ilari Maasilta, Henk Hoevers, Sae Woo Nam, Jonas Zmuidzinas, Harvey Moseley and Ben Mazin. PJdV is financially supported by the Netherlands Organisation for Scientific Research NWO (Veni Grant No. 639.041.750 and Projectruimte 680-91-127). JJAB was supported by the European Research Counsel ERC (Consolidator Grant No. 648135 MOSAIC). 

\appendix

\section{Phonon trapping model}\label{app:phonontrapping}
The phonon trapping we rely on in the membrane experiment is purely geometrical: the Al film is much wider ($1.7~\mu$m) than the membrane thickness (110 nm), which leads to reflection of the phonons back into the Al film. To estimate the effective phonon trapping due to the thin membrane underneath the superconducting strip, we perform a simulation with a geometrical ray tracing model. In the downconversion process from photon energy to quasiparticles, phonons play a prominent role. For this problem we are mainly interested in phonons with relatively high energies and therefore short phonon wavelengths. Phonons with energies ranging from the Debye energy (37 meV in Al) to around 10$\Delta$ = 1.9 meV dominate this process \cite{akozorezov2000}, which corresponds to typical phonon wavelengths of $\sim$0.5--10 nm. These wavelengths are small enough to expect that phonon dimensionality does not play a role yet and a geometrical model is a reasonable assumption. 

In the model, we initialise a phonon at the top surface of the metal film at a random position and with a random angle of propagation and trace the propagation of the phonon. Reflection at any interface is assumed to randomise the angle of reflection due to roughness, in comparison to the short wavelengths. When hitting the metal-substrate interface we calculate the probability of the transmission from the acoustic mismatch model with the equations from Ref. \onlinecite{skaplan1979}. Upon transmission the phonon is refracted according to Snell's law. 

When the phonon enters the area underneath the film we distinguish between the substrate and membrane. For a film on the substrate, we consider a phonon that enters the substrate to be lost, since the substate is 350 $\mu$m thick, compared to the metal strip of 1.7 $\mu$m wide. 

For the membrane, a phonon that hits the bottom of the membrane is reflected back at random angle, and therefore has a probability to re-enter the film, again with a probability from the acoustic matching. When the phonon propagates outside the area underneath the metal, we estimate in this case a 50\% probability of returning underneath the film at a random angle, to take into account that any (diffusive) interaction with the membrane interfaces outside of this area can lead to the phonon returning. Diffusive scattering at the membrane interfaces is plausible considering that the membrane is exposed to the reactive ion etch for the NbTiN on the front side. Phonons that do not return we consider lost.

\begin{table*}[t]
\centering
\begin{tabular}{ccccccc}
\hline
Material	 & Thickness				& $c_t$ (km/s)	& $c_l$ (km/s)	& density (g/cm$^3$) 	& $\tau_{esc}$ (ns) & $\tau_{esc}$ (ns) \\ 
	& & & & & 50\% return & no return \\
\hline
Al		&  50 nm					& 3.26		&  6.65		& 2.73				& n.a.		& n.a.	\\ 	
Sapphire	&  350 $\mu$m				& 6.45	 	& 10.9		& 3.99				& n.a.		& 0.18	\\ 	
SiN		&  350 $\mu$m 			& 6.2			& 10.3		& 3.1 				& n.a.		& 0.16	\\   	
SiN		&  110 nm					& 6.2			& 10.3		& 3.1 				& 1.3			& 0.74 	\\	
\hline
\end{tabular}
\caption{Overview of the phonon properties used in the ray tracing model and the resulting phonon escape times from an Al film on top of that substrate. In the model we take the 110 nm SiN on Si substrate as 350 $\mu$m SiN. Properties of Al and sapphire are taken from Ref. \onlinecite{skaplan1979} and SiN from Refs. \onlinecite{tkuhn2004,kpetersen1982}.}\label{table:phonontrapping}
\end{table*}

The effect of phonon trapping is quantified using the total dwell time of the phonon in the superconducting film, usually called the phonon escape time $\tau_{esc}$. We average the dwell time of 100,000 phonons for each set of parameters. The results are shown in Table \ref{table:phonontrapping}. For an Al film on a 350 $\mu$m thick sapphire substrate using this model we find reasonable agreement with the analytical expression $\tau_{esc} = 4d/\eta u$, with $u$ the average phonon velocity and $\eta$ the interface transparency (when integrated over all angles), using numbers from Ref. \onlinecite{skaplan1979}. 

Comparing the dwell times on SiN substrate and SiN membrane, we find a factor of 8.1 enhancement due to the thin membrane. This corresponds very well to the trapping factor measured from resolving power $8\pm2$, especially given the simplicity of the model.

The phonon trapping enhancement obtained from the measurements of quasiparticle fluctuations at temperatures from $250-350$ mK is $66\pm17$ \cite{sderooij2021}. The equilibrium generation-recombination process involves only phonons with energies close to $2\Delta$. An energy of $2\Delta$ corresponds to a phonon wavelength of $\sim$50 nm, which is comparable to the thicknesses of film and membrane. Therefore one would not a-priori expect a close agreement between the effective trapping at high energies and these low energies. A complete model of phonon propagation in a thin membrane at the different energies involved in the downconversion process (from the Debye energy down to $2\Delta$) would be needed to microscopically model the phonon propagation. This model should include the preferential phonon modes emitted from the aluminium film, the density of phonon states of film and membrane and the phonon scattering at the (etched) surfaces of the membrane.

\section{Low energy events for membrane devices}\label{app:lowenergyhits}

\begin{figure}
\includegraphics[width=0.99\columnwidth]{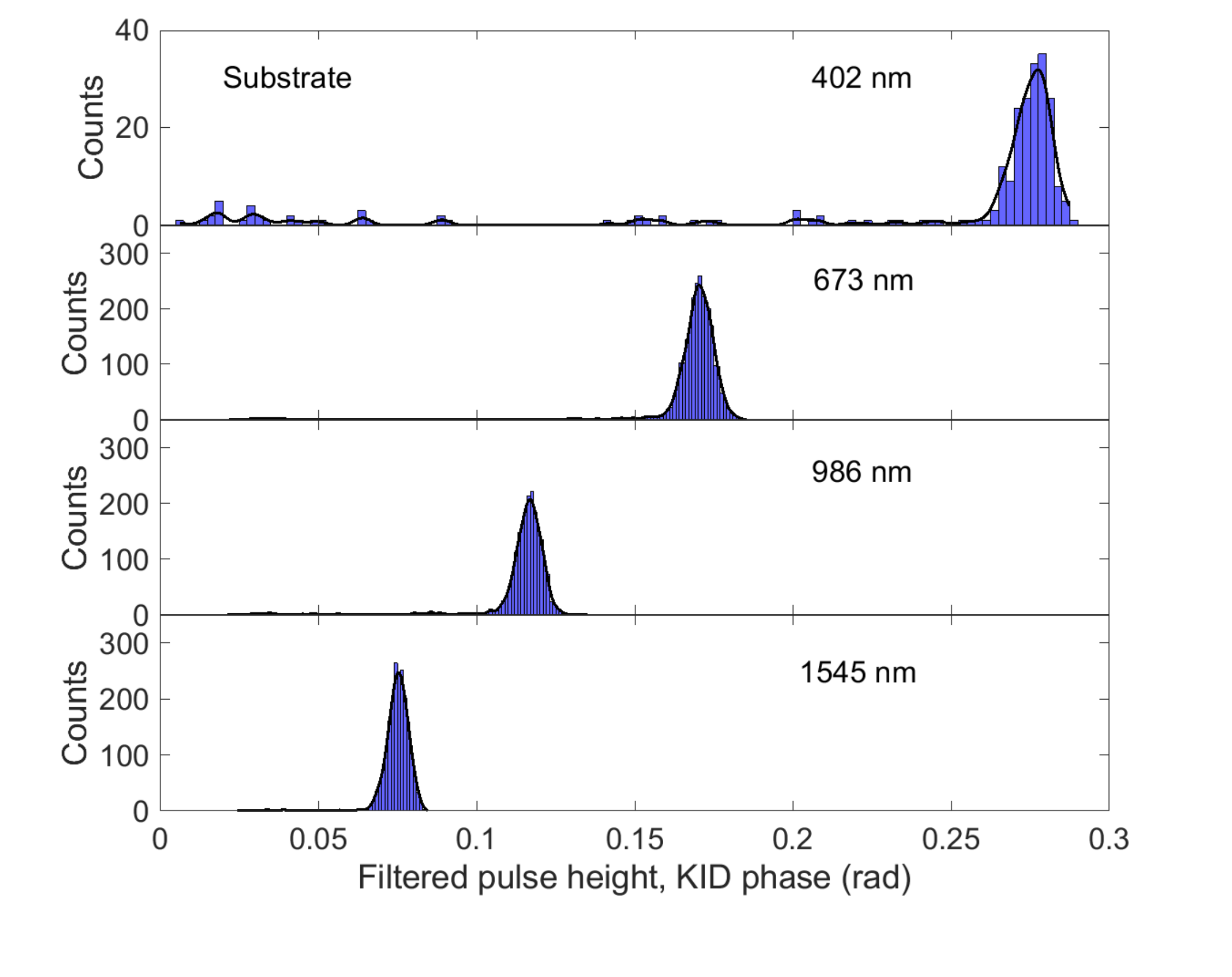}
\caption{\label{fig:histogramsKID3} Full histograms for the laser wavelengths as indicated for Device C, at -83 dBm and 120 mK. The resolving powers following from the kernel density estimate (lines) are presented in Fig. \ref{fig:resolvingpower}a.}
\end{figure}

\begin{figure}
\includegraphics[width=0.99\columnwidth]{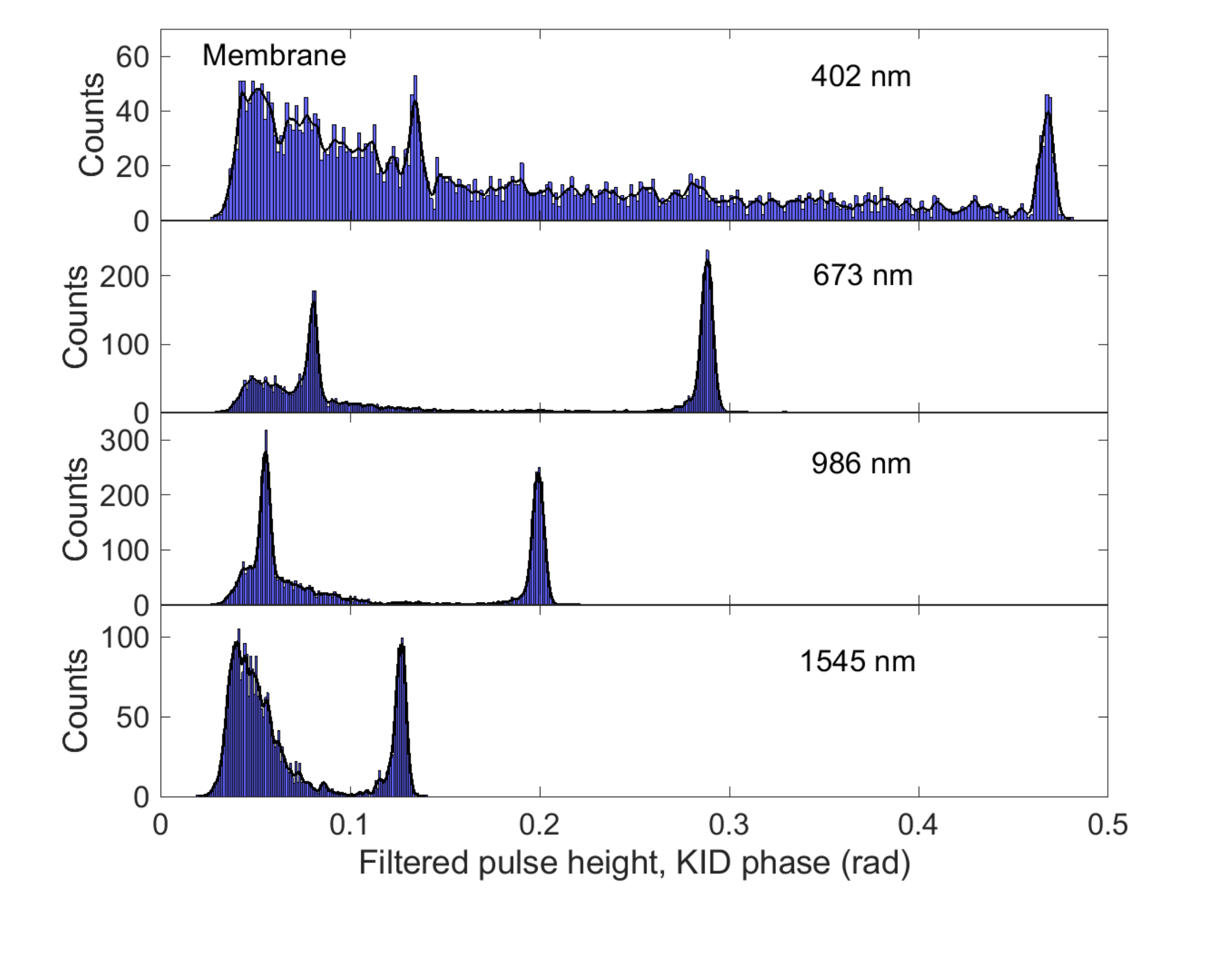}
\caption{\label{fig:histogramsKID2} Full histograms for the laser wavelengths as indicated for Device D, at -85 dBm and 120 mK. The resolving powers following from the kernel density estimate (lines) are presented in Fig. \ref{fig:resolvingpower}a.}
\end{figure}

The full histograms of Devices C and D, corresponding to the resolving power measurements in Fig. \ref{fig:resolvingpower}a, are shown in Figures \ref{fig:histogramsKID3} and \ref{fig:histogramsKID2} respectively. Comparing the substrate and membrane based MKIDs, we find that low energy hits are more frequent in the membrane-based device D. In the membrane device there is typically a strong peak in the histogram at the expected pulse height for direct Al hits (at the highest pulse height). Then there is broad shoulder at low energies, most likely related to NbTiN hits, since this shoulder is much stronger for 402 nm light, where NbTiN is more absorptive. The third class of hits is represented by a peak at low energy, with typically $\sim$30\% the pulse height of the primary hits. For the substrate MKIDs, the signal-to-noise is also high enough to see both classes of low energy hits if they would occur, but this is not the case (Fig. \ref{fig:histogramsKID3}). On top of that, the number of direct Al hits per unit time is similar for devices C and D for the same laser power. We therefore attribute both classes of low energy hits to photons being absorbed in other parts of the detector and leading to a response through phonon transport in the membrane. The broad shoulder in the histogram can therefore be attributed to the large NbTiN areas on the left and right of the membrane (see Fig. \ref{fig:designs}e). These hits lead to different pulse heights because energy can leak out due to diffusion and phonon transport into the substrate. The low energy peak is most likely related to the NbTiN connections in the centre of the membrane where it is more likely that a well-defined amount of energy reaches the Al. The low energy hits show up in the histogram far enough from the primary peak to not negatively affect the measured $R$, which we take as the FWHM of the primary peak.

The low-energy events could be avoided by properly focusing the radiation onto the sensitive part of the MKID with a microlens. As explained in Section \ref{sec:kiddesign} we have chosen a well-studied geometry with sufficient signal-to-noise for energy resolution studies, with the downside of non-selective illumination. 

\begin{figure}
\includegraphics[width=0.99\columnwidth]{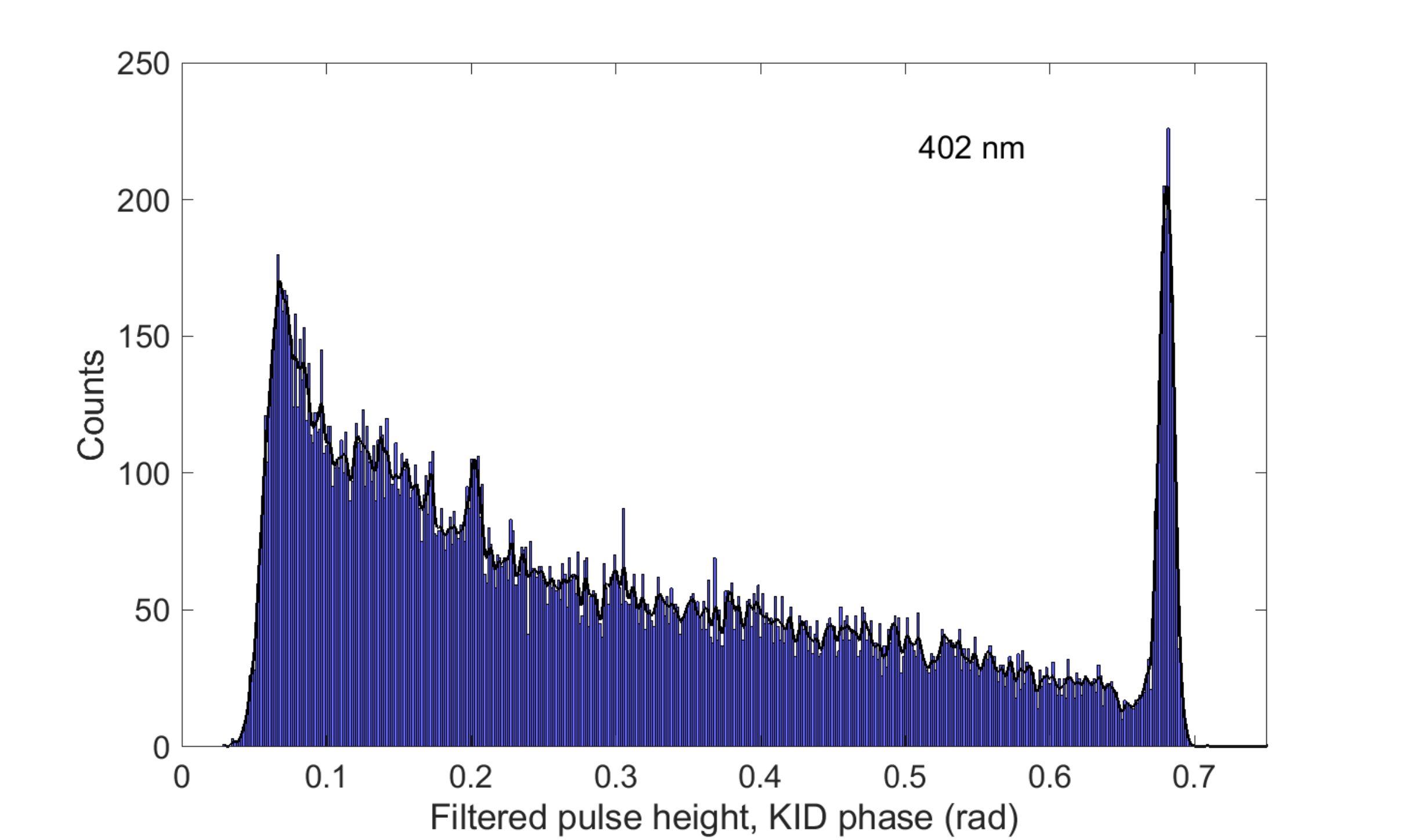}
\caption{\label{fig:longhistogramKID2} Full histogram for the 402 nm laser for Device D, at -87 dBm and 120 mK, based on 26306 counts (3987 in Fig. \ref{fig:histogramsKID2}). The resulting resolving power is the same as in Fig. \ref{fig:histogramsKID2}.}
\end{figure}

For Device D at 402 nm, as shown in the top panel of Fig. \ref{fig:histogramsKID2}, we also measured a longer timetrace (1024 s) in a separate cooldown to check the resolving power with more counts in the histogram, which is shown in Fig. \ref{fig:longhistogramKID2}, and gives a consistent $R$ with Fig. \ref{fig:histogramsKID2}. The average pulseheight is higher than in Fig. \ref{fig:histogramsKID2}, because the coupling quality factor changes from cooldown to cooldown, here from 48.5$\times 10^3$ to 77.7$\times 10^3$, most likely due to the very open IDC structure.

Because of the high signal-to-noise, subsequent single photon hits that are too close in time are generally easy to distinguish and filter out before making the histogram. However, in the presence of low-energy hits on device D this is not always effective. We therefore need to remove double hits that involve a primary and a low-energy hit in a post-processing step, for which we reject these double hits based on the residual after applying the optimal filter, $\theta(f) - AM(f)$, integrated over frequency (see Fig. \ref{fig:optimalfilter}b). We have verified that this step only removes double hits and does not artificially reduce the number of low-enery hits, which is evident from Figures \ref{fig:histogramsKID2} and \ref{fig:longhistogramKID2}.\\

\section{Equations for the low-energy part of the downconversion}\label{app:eqdownconversion}
The equations for the low energy-part of the downconversion, $J_{low}$, are taken from Ref. \onlinecite{akozorezov2008} (Eq. 6 therein). We reproduce them here in the original notation with corrections\cite{akozorezov_comm}.

\begin{eqnarray}
J_{low} = \eta(1-\xi_c^2)\frac{\Omega_D}{\epsilon_0}\frac{l_{pe,D}}{d} \frac{3 Z_1(0)}{11Z_1(0)+3} g_1\left(\frac{\Omega_D}{\Omega_1}\right),
\label{eq:Jlow}
\end{eqnarray}
Here we take $Z_1(0)=1$. $g_1(x)$ is given by:
\begin{widetext}
\begin{eqnarray}
\begin{split}
g_1(x) = \int^1_{1/x} dz f(z) \bigg\{&Ei(1,z(x-1))-Ei(1,1-z) +\ln\left(\frac{(x-1)z}{1-z}\right)\\
 &-\frac{1}{4}\eta(1-\xi_c)\left[Ei(1,2z(x-1))-Ei(1,2(1-z)) +\ln\left(\frac{(x-1)z}{1-z}\right)\right] \bigg\},
\end{split}
\label{eq:g1}
\end{eqnarray}
where the $\ln(...)$ terms are corrected with respect to Ref. \onlinecite{akozorezov2008}. $f(x)$ is given by
\begin{equation*}
f(x) = 1-x-\frac{1}{12}x\left[x^2\left(\cos\left(\sqrt{2}\ln x\right)-7\sqrt{2}\sin\left(\sqrt{2}\ln x\right)\right)-1\right].
\label{eq:f}
\end{equation*}
\end{widetext}
Here $\Omega_D$ is the Debye energy, $\Omega_1$ is the characteristic energy at which this stage ends, $\eta$ is the acoustical transmission coefficient, $l_{pe,D}$ is the mean free path, $d$ the film thickness, $\epsilon_0=1.7\Delta$ the average quasiparticle energy and $\xi_c = \cos(\theta_c) $, with $\theta_c$ the critical angle for phonon transmission at the interface.

%

\end{document}